\documentclass[letterpaper]{acmproc}
\usepackage[utf8]{inputenc}
\usepackage{color}
\usepackage[setpagesize=false]{hyperref}
\usepackage{epsfig}

\usepackage{microtype}
\usepackage{balance}

\newdimen\oldframetabcolsep
\newdimen\oldcolortabcolsep
\newdimen\oldpretabcolsep

\global\boilerplate{Copyright \copyright{} 2013 Ludovic Courtès

  Permission is granted to copy, distribute and/or modify this document
  under the terms of the GNU Free Documentation License, Version 1.3
  or any later version published by the Free Software Foundation;
  with no Invariant Sections, no Front-Cover Texts, and no Back-Cover Texts.
  A copy of the license is
  available at \url{http://www.gnu.org/licenses/gfdl.html}.

  The source of this document is available from
  \url{http://git.sv.gnu.org/cgit/guix/maintenance.git}.
}
\conferenceinfo{European Lisp Symposium}{2013, Madrid, Spain}

\title{Functional Package Management with Guix}
\numberofauthors{1}
\author{
\alignauthor
Ludovic Court\`{e}s\\
Bordeaux, France\\
ludo@gnu.org\\
}
\begin{document}
\date{}
\maketitle
\clubpenalty=500
\widowpenalty=500

\subsection*{ABSTRACT}
\noindent \noindent We describe the design and implementation of GNU~Guix, a %
purely functional package manager designed to support a complete %
GNU/Linux distribution.  Guix supports transactional upgrades and %
roll-backs, unprivileged package management, per-user profiles, and %
garbage collection.  It builds upon the low-level build and deployment %
layer of the Nix package manager.  Guix uses Scheme as its programming %
interface.  In particular, we devise an embedded domain-specific %
language (EDSL) to describe and compose packages.  We demonstrate how it %
allows us to benefit from the host general-purpose programming language %
while not compromising on expressiveness.  Second, we show the use of %
Scheme to write build programs, leading to a ``two-tier'' programming %
system.\par
\par
\terms{Languages, Management, Reliability}\keywords{Functional package management, Scheme, Embedded %
domain-specific language}\newpage

\section{Introduction}
\label{chapter8267}
\noindent GNU~Guix\footnote{\href{http://www.gnu.org/software/guix/}{{\texttt{http{\char58}/\-/\-www.\-gnu.\-org/\-software/\-guix/\-}}}} is a {\em{purely functional}} %
package manager for the GNU system [20], %
and in particular GNU/Linux.  Package management consists in all the %
activities that relate to building packages from source, honoring the %
build-time and run-time dependencies on packages, installing, removing, %
and upgrading packages in user environments.  In addition to these %
standard features, Guix supports transactional upgrades and roll-backs, %
unprivileged package management, per-user profiles, and garbage %
collection.  Guix comes with a distribution of user-land free software %
packages.\par
\noindent Guix seeks to empower users in several ways{\char58} by offering the %
uncommon features listed above, by providing the tools that allow users %
to formally correlate a binary package and the ``recipes'' and source %
code that led to it---furthering the spirit of the GNU~General %
Public License---, by allowing them to customize the distribution, %
and by lowering the barrier to entry in distribution development.\par
\noindent The keys toward these goals are the implementation of a %
purely functional package management paradigm, and the use of both %
declarative and lower-level programming interfaces (APIs) embedded in %
Scheme.  To that end, Guix reuses the package storage and deployment %
model implemented by the Nix functional package manager [8].  On top of that, it provides Scheme APIs, and in %
particular embedded domain-specific languages (EDSLs) to describe %
software packages and their build system.  Guix also uses Scheme for %
programs and libraries that implement the actual package build %
processes, leading to a ``two-tier'' system.\par
\noindent This paper focuses on the programming techniques implemented by %
Guix.  Our contribution is twofold{\char58} we demonstrate that use of Scheme %
and EDSLs achieves expressiveness comparable to that of Nix's DSL while %
providing a richer and extensible programming environment; we further %
show that Scheme is a profitable alternative to shell tools when it %
comes to package build programs.  \hyperref[background]{Section~2} first gives some background on functional package %
management and its implementation in Nix.  \hyperref[api]{Section~3} describes the design and implementation of Guix's %
programming and packaging interfaces.  \hyperref[eval]{Section~4} provides an evaluation and discussion of the current status of %
Guix.  \hyperref[related]{Section~5} presents related work, %
and \hyperref[conclusion]{Section~6} concludes.\par
\newpage

\section{Background}
\label{background}
\noindent This section describes the functional package management %
paradigm and its implementation in Nix.  It then shows how Guix differs, %
and what the rationale is.\par

\subsection{Functional Package Management}
\label{section8290}
\noindent Functional package management is a paradigm whereby the %
build and installation process of a package is considered as a pure %
function, without any side effects.  This is in contrast with %
widespread approaches to package build and installation where the build %
process usually has access to all the software installed on the machine, %
regardless of what its declared inputs are, and where installation %
modifies files in place.\par
\noindent Functional package management was pioneered by the Nix %
package manager [8], which has since matured to %
the point of managing a complete GNU/Linux distribution [9].  To allow build processes to be faithfully %
regarded as pure functions, Nix can run them in a {\texttt{chroot}} %
environment that only contains the inputs it explicitly declared; thus, %
it becomes impossible for a build process to use, say, Perl, if that %
package was not explicitly declared as an input of the build process. %
In addition, Nix maps the list of inputs of a build process to a %
statistically unique file system name; that file name is used to %
identify the output of the %
build process.  For instance, a particular build of GNU~Emacs may be %
installed in {\texttt{/\-nix/\-store/\-v9zic07iar8w90zcy398r745w78a7lqs-emacs-24.\-2}}, based on a %
cryptographic hash of all the inputs to that build process; changing the %
compiler, configuration options, build scripts, or any other inputs to %
the build process of Emacs yields a different name.  This is a %
form of {\em{on-disk memoization}}, with the {\texttt{/\-nix/\-store}} %
directory acting as a cache of ``function results''---i.e., a cache %
of installed packages.  Directories under {\texttt{/\-nix/\-store}} are %
immutable.\par
\noindent This direct mapping from build inputs to the result's %
directory name is basis of the most important properties of a functional package %
manager.  It means that build processes are regarded as {\em{referentially transparent}}.  To put it differently, instead %
of merely providing pre-built binaries %
and/or build recipes, functional package managers provide binaries, %
build recipes, and in effect a {\em{guarantee}} that a given binary %
matches a given build recipe.\par

\subsection{Nix}
\label{section8356}
\noindent The idea of {\em{purely functional}} package started by %
making an analogy between programming language paradigms and software %
deployment techniques [8].  The authors %
observed that, in essence, package management tools typically used on %
free operating systems, such as RPM and Debian's APT, implement an %
{\em{imperative}} software deployment paradigm.  Package %
installation, removal, and upgrade are all done in-place, by mutating %
the operating system's state.  Likewise, changes to the operating %
system's configuration are done in-place by changing configuration %
files.\par
\noindent This imperative approach has several drawbacks.  First, it %
makes it hard to reproduce or otherwise describe the OS state.  Knowing %
the list of installed packages and their version is not enough, because %
the installation procedure of packages may trigger hooks to change %
global system configuration files [4, 7], and of course users may have done %
additional modifications.  Second, installation, removal, and upgrade %
are not transactional; interrupting them may leave the system in an %
undefined, or even unusable state, where some of the files have been %
altered.  Third, rolling back to a previous system configuration is %
practically impossible, due to the absence of a mechanism to formally %
describe the system's configuration.\par
\noindent Nix attempts to address these shortcomings through the %
functional software deployment paradigm{\char58} installed packages are %
immutable, and build processes are regarded as pure functions, as %
explained before.  Thanks to this property, it implements {\em{transparent source/binary deployment}}{\char58} the directory name of a build %
result encodes all the inputs of its build process, so if a trusted %
server provides that directory, then it can be directly downloaded from %
there, avoiding the need for a local build.\par
\noindent Each user has their own {\em{profile}}, which contains %
symbolic links to the {\texttt{/\-nix/\-store}} directories of installed %
packages.  Thus, users can install packages independently, and the %
actual storage is shared when several users install the very same %
package in their profile.  Nix comes with a {\em{garbage collector}}, %
which has two main functions{\char58} with conservative scanning, it can %
determine what packages a build output refers to; and upon user request, %
it can delete any packages not referenced {\textit{via}} any user %
profile.\par
\noindent To describe and compose build processes, Nix implements its %
own domain-specific language (DSL), which provides a convenient %
interface to the build and storage mechanisms described above.  The Nix %
language is purely functional, lazy, and dynamically typed; it is similar %
to that of the Vesta software configuration system [11].  It comes %
with a handful of built-in data types, and around 50 primitives.  The %
primitive to describe a build process is {\texttt{derivation}}.\par
\begin{figure}[ht]
\setlength{\oldpretabcolsep}{\tabcolsep}
\addtolength{\tabcolsep}{-\tabcolsep}
{\setbox1 \vbox \bgroup
{\noindent \texttt{\begin{tabular}{l}
{\textit{\ \ 1{\char58}\ }}\ \ derivation\ \{\\
{\textit{\ \ 2{\char58}\ }}\ \ \ \ name\ =\ "example-1.0";\\
{\textit{\ \ 3{\char58}\ }}\ \ \ \ builder\ =\ "\$\{./static-bash\}";\ \ \ \ \ \ \ \ \ \ \ \ \ \ \ \ \ \\
{\textit{\ \ 4{\char58}\ }}\ \ \ \ args\ =\ {\char91}\ "-c"\ "echo\ hello\ $>$\ \$out"\ {\char93};\\
{\textit{\ \ 5{\char58}\ }}\ \ \ \ system\ =\ "x86\_64-linux";\\
{\textit{\ \ 6{\char58}\ }}\ \ \}\\
\end{tabular}
}}
\egroup{\box1}}%
\setlength{\tabcolsep}{\oldpretabcolsep}
\caption{\label{fig:derivation-nix}Call to the {\texttt{derivation}} primitive in %
the Nix language.}\end{figure}
\noindent Figure~\ref{fig:derivation-nix} shows code that calls %
the {\texttt{derivation}} function with one argument, which is a %
dictionary.  It expects at least the four key/value pairs shown above; %
together, they define the build process and its inputs.  The result is a %
{\em{derivation}}, which is essentially the {\em{promise}} of a %
build.  The derivation has a low-level on-disk representation %
independent of the Nix language---in other words, derivations are to %
the Nix language what assembly is to higher-level programming %
languages.  When %
this derivation is instantiated---i.e., built---, it runs the command %
{\texttt{static-bash -c "echo hello $>$ \$out"}} in a chroot that contains %
nothing but the {\texttt{static-bash}} file; in addition, each key/value %
pair of the {\texttt{derivation}} argument is reified in the build process %
as an environment variable, and the {\texttt{out}} environment variable is %
defined to point to the output {\texttt{/\-nix/\-store}} file name.\par
\noindent Before the build starts, the file {\texttt{static-bash}} is %
imported under {\texttt{/\-nix/\-store/\-}}...{\texttt{-static-bash}}, and the value %
associated with {\texttt{builder}} is substituted with that file name.  This {\texttt{\$\{.\-.\-.\-\}}} form on line 3 for string interpolation makes it easy to insert %
Nix-language values, and in particular computed file names, in the %
contents of build scripts.  The Nix-based GNU/Linux distribution, NixOS, %
has most of its build scripts written in Bash, and makes %
heavy use of string interpolation on the Nix-language side.\par
\noindent All the files referenced by derivations live under {\texttt{/\-nix/\-store}}, called {\em{the store}}.  In a multi-user setup, users %
have read-only access to the store, and all other accesses to the store %
are mediated by a daemon running as {\texttt{root}}.  Operations such as %
importing files in the store, computing a derivation, building a %
derivation, or running the garbage collector are all implemented as %
remote procedure calls (RPCs) to the daemon.  This guarantees that the %
store is kept in a consistent state---e.g., that referenced files and %
directories are not garbage-collected, and that the contents of files %
and directories are genuine build results of the inputs hashed in their %
name.\par
\noindent The implementation of the Nix language is an interpreter %
written in C++.  In terms of performance, it does not compete with %
typical general-purpose language implementations; that is often not a %
problem given its specific use case, but sometimes requires rewriting %
functions, such as list-processing tools, as language primitives in %
C++.  The language itself is not extensible{\char58} it has no macros, a fixed %
set of data types, and no foreign function interface.\par

\subsection{From Nix to Guix}
\label{nix-to-guix}
\noindent Our main contribution with GNU~Guix is the use of Scheme %
for both the composition and description of build processes, and the %
implementation of build scripts.  In other words, Guix builds upon the %
build and deployment primitives of Nix, but replaces the Nix language by %
Scheme with embedded domain-specific languages (EDSLs), and promotes %
Scheme as a replacement for Bash in build scripts.  Guix is implemented %
using GNU~Guile~2.0\footnote{\href{http://www.gnu.org/software/guile/}{{\texttt{http{\char58}/\-/\-www.\-gnu.\-org/\-software/\-guile/\-}}}}, a rich implementation of Scheme %
based on a compiler and bytecode interpreter that supports the R5RS and %
R6RS standards.  It reuses the build primitives of Nix by making remote %
procedure calls (RPCs) to the Nix build daemon.\par
\noindent We claim that using an {\em{embedded}} DSL has numerous %
practical benefits over an independent DSL{\char58} tooling (use of Guile's %
compiler, debugger, and REPL, Unicode support, etc.), libraries (SRFIs, %
internationalization support, etc.), and seamless integration in larger %
programs.  To illustrate this last point, consider an application that %
traverses the list of available packages and processes it---for instance %
to filter packages whose name matches a pattern, or to render it as %
HTML.  A Scheme program can readily and efficiently do it with Guix, %
where packages are first-class Scheme objects; conversely, writing such %
an implementation with an external DSL such as Nix requires either %
extending the language implementation with the necessary functionality, %
or interfacing with it {\textit{via}} an external representation such as %
XML, which is often inefficient and lossy.\par
\noindent We show that use of Scheme in build scripts is natural, and %
can achieve conciseness comparable to that of shell scripts, but with %
improved expressivity and clearer semantics.\par
\noindent The next section describes the main programming interfaces %
of Guix, with a focus on its high-level package description language and %
``shell programming'' substitutes provided to builder-side code.\par

\section{Build Expressions and Package Descriptions}
\label{api}
\noindent Our goal when designing Guix was to provide interfaces %
ranging from Nix's low-level primitives such as {\texttt{derivation}} to high-level package declarations.  The declarative %
interface is a requirement to help grow and maintain a large software %
distribution.  This section describes the three level of abstractions %
implemented in Guix, and illustrates how Scheme's homoiconicity and %
extensibility were instrumental.\par

\subsection{Low-Level Store Operations}
\label{section8410}
\noindent As seen above, {\em{derivations}} are the central %
concept in Nix.  A derivation bundles together a {\em{builder}} and %
its execution environment{\char58} command-line arguments, environment variable %
definitions, as well as a list of input derivations whose result should %
be accessible to the builder.  Builders are typically executed in a {\texttt{chroot}} environment where only those inputs explicitly listed are %
visible.  Guix transposes Nix's {\texttt{derivation}} primitive literally %
to its Scheme interface.\par
\begin{figure}[ht]
\setlength{\oldpretabcolsep}{\tabcolsep}
\addtolength{\tabcolsep}{-\tabcolsep}
{\setbox1 \vbox \bgroup
{\noindent \texttt{\begin{tabular}{l}
{\textit{\ \ 1{\char58}\ }}(let*\ ((store\ (open-connection))\ \ \ \ \ \ \ \ \ \ \ \ \ \ \ \ \ \ \ \ \ \\
{\textit{\ \ 2{\char58}\ }}\ \ \ \ \ \ \ (bash\ \ (add-to-store\ store\ "static-bash"\ \ \ \ \ \ \\
{\textit{\ \ 3{\char58}\ }}\ \ \ \ \ \ \ \ \ \ \ \ \ \ \ \ \ \ \ \ \ \ \ \ \ \ \ \ \#t\ "sha256"\\
{\textit{\ \ 4{\char58}\ }}\ \ \ \ \ \ \ \ \ \ \ \ \ \ \ \ \ \ \ \ \ \ \ \ \ \ \ \ "./static-bash")))\\
{\textit{\ \ 5{\char58}\ }}\ \ (derivation\ store\ "example-1.0"\ \\
{\textit{\ \ 6{\char58}\ }}\ \ \ \ \ \ \ \ \ \ \ \ \ \ "x86\_64-linux"\ \\
{\textit{\ \ 7{\char58}\ }}\ \ \ \ \ \ \ \ \ \ \ \ \ \ bash\\
{\textit{\ \ 8{\char58}\ }}\ \ \ \ \ \ \ \ \ \ \ \ \ \ '("-c"\ "echo\ hello\ $>$\ \$out")\\
{\textit{\ \ 9{\char58}\ }}\ \ \ \ \ \ \ \ \ \ \ \ \ \ '()\ '()))\ \ \ \ \ \ \ \ \ \ \ \ \ \ \ \ \ \ \ \ \ \ \ \ \ \ \ \ \ \ \\
{\textit{\ 10{\char58}\ }}\\
{\textit{\ 11{\char58}\ }}\begin{math}\Rightarrow\end{math}\ \ \ \ \ \ \ \ \ \ \ \ \ \ \ \ \ \ \ \ \ \ \ \ \ \ \ \ \ \ \ \ \ \ \ \ \ \ \ \ \ \ \ \ \ \ \ \ \\
{\textit{\ 12{\char58}\ }}"/nix/store/nsswy...-example-1.0.drv"\\
{\textit{\ 13{\char58}\ }}\#$<$derivation\ "example-1.0"\ ...$>$\\
\end{tabular}
}}
\egroup{\box1}}%
\setlength{\tabcolsep}{\oldpretabcolsep}
\caption{\label{fig:derivation-prim}Using the {\texttt{derivation}} primitive in %
Scheme with Guix.}\end{figure}
\noindent Figure~\ref{fig:derivation-prim} shows the %
example of Figure~\ref{fig:derivation-nix} rewritten to use %
Guix's low-level Scheme API.  Notice how the former makes explicit %
several operations not visible in the latter.  First, line 1 establishes a connection to the build daemon; line 2 explicitly asks the daemon to ``intern'' file {\texttt{static-bash}} into the store; finally, the {\texttt{derivation}} call %
instructs the daemon to compute the given derivation.  The two arguments %
on line~9 are a set of environment variable %
definitions to be set in the build environment (here, it's just the %
empty list), and a set of {\em{inputs}}---other derivations %
depended on, and whose result must be available to the build process. %
Two values are %
returned (line 11){\char58} the file name of the on-disk %
representation of the derivation, and its in-memory representation as a %
Scheme record.\par
\noindent The build actions represented by this derivation can then %
be performed by passing it to the {\texttt{build-derivations}} RPC.  Again, %
its build result is a single file reading {\texttt{hello}}, and its build is %
performed in an environment where the only visible file is a copy of %
{\texttt{static-bash}} under {\texttt{/\-nix/\-store}}.\par

\subsection{Build Expressions}
\label{build-exprs}
\noindent The Nix language heavily relies on string interpolation to %
allow users to insert references to build results, while hiding the %
underlying {\texttt{add-to-store}} or {\texttt{build-derivations}} operations %
that appear explicitly in Figure~\ref{fig:derivation-prim}.  Scheme %
has no support for string interpolation; adding it to the underlying %
Scheme implementation is certainly feasible, but it's also unnatural.\par
\noindent The obvious strategy here is to instead leverage Scheme's %
homoiconicity.  This leads us to the definition of {\texttt{build-expression-$>$derivation}}, which works similarly to {\texttt{derivation}}, except that it expects a {\em{build expression}} as an %
S-expression instead of a builder.  Figure~\ref{fig:expr-derivation} shows the same derivation as before but rewritten %
to use this new interface.\par
\begin{figure}[ht]
\setlength{\oldpretabcolsep}{\tabcolsep}
\addtolength{\tabcolsep}{-\tabcolsep}
{\setbox1 \vbox \bgroup
{\noindent \texttt{\begin{tabular}{l}
{\textit{\ \ 1{\char58}\ }}(let\ ((store\ \ \ (open-connection))\\
{\textit{\ \ 2{\char58}\ }}\ \ \ \ \ \ (builder\ '(call-with-output-file\ \%output\ \ \ \ \ \\
{\textit{\ \ 3{\char58}\ }}\ \ \ \ \ \ \ \ \ \ \ \ \ \ \ \ \ \ (lambda\ ()\\
{\textit{\ \ 4{\char58}\ }}\ \ \ \ \ \ \ \ \ \ \ \ \ \ \ \ \ \ \ \ (display\ "hello")))))\\
{\textit{\ \ 5{\char58}\ }}\ \ (build-expression-$>$derivation\ store\ \\
{\textit{\ \ 6{\char58}\ }}\ \ \ \ \ \ \ \ \ \ \ \ \ \ \ \ \ \ \ \ \ \ \ \ \ \ \ \ \ \ \ \ "example-1.0"\\
{\textit{\ \ 7{\char58}\ }}\ \ \ \ \ \ \ \ \ \ \ \ \ \ \ \ \ \ \ \ \ \ \ \ \ \ \ \ \ \ \ \ "x86\_64-linux"\\
{\textit{\ \ 8{\char58}\ }}\ \ \ \ \ \ \ \ \ \ \ \ \ \ \ \ \ \ \ \ \ \ \ \ \ \ \ \ \ \ \ \ builder\ '()))\\
{\textit{\ \ 9{\char58}\ }}\\
{\textit{\ 10{\char58}\ }}\begin{math}\Rightarrow\end{math}\\
{\textit{\ 11{\char58}\ }}"/nix/store/zv3b3...-example-1.0.drv"\\
{\textit{\ 12{\char58}\ }}\#$<$derivation\ "example-1.0"\ ...$>$\\
\end{tabular}
}}
\egroup{\box1}}%
\setlength{\tabcolsep}{\oldpretabcolsep}
\caption{\label{fig:expr-derivation}Build expression written in Scheme.}\end{figure}
\noindent This time the builder on line 2 is %
purely a Scheme expression.  That expression will be evaluated when the %
derivation is built, in the specified build environment with no inputs. %
The environment implicitly includes a copy of Guile, which is used to %
evaluate the {\texttt{builder}} expression.  By default this is a %
stand-alone, statically-linked Guile, but users can also specify a %
derivation denoting a different Guile variant.\par
\noindent Remember that this expression is run by a separate Guile %
process than the one that calls {\texttt{build-expression-$>$derivation}}{\char58} it %
is run by a Guile process launched by the build daemon, in a {\texttt{chroot}}.  So, while there is a single language for both the ``host'' %
and the ``build'' side, there are really two {\em{strata}} of code, %
or {\em{tiers}}{\char58} the host-side, and the build-side code\footnote{The term ``stratum'' is this context was coined by Manuel Serrano et %
al. for their work on Hop where a similar situation arises [17].}.\par
\noindent Notice how the output file name is reified {\textit{via}} the %
{\texttt{\%output}} variable automatically added to {\texttt{builder}}'s scope. %
Input file names are similarly reified through the {\texttt{\%build-inputs}} %
variable (not shown here).  Both variables are non-hygienically %
introduced in the build expression by {\texttt{build-expression-$>$derivation}}.\par
\noindent Sometimes the build expression needs to use functionality %
from other modules.  For modules that come with Guile, the expression %
just needs to be augmented with the needed {\texttt{(use-modules .\-.\-.\-)}} %
clause.  Conversely, external modules first need to be imported into the %
derivation's build environment so the build expression can use them.  To %
that end, the {\texttt{build-expression-$>$derivation}} procedure has an %
optional {\texttt{\#{\char58}modules}} keyword parameter, allowing additional %
modules to be imported into the expression's environment.\par
\noindent When {\texttt{\#{\char58}modules}} specifies a non-empty module list, an %
auxiliary derivation is created and added as an input to the initial %
derivation.  That auxiliary derivation %
copies the module source and compiled files in the store.  This %
mechanism allows build expressions to easily use helper modules, as %
described in \hyperref[shell]{Section~3.4}.\par

\subsection{Package Declarations}
\label{pkg-decl}
\begin{figure*}[!th]
\setlength{\oldpretabcolsep}{\tabcolsep}
\addtolength{\tabcolsep}{-\tabcolsep}
{\setbox1 \vbox \bgroup
{\noindent \texttt{\begin{tabular}{l}
{\textit{\ \ 1{\char58}\ }}({\textcolor[rgb]{0.4117647058823529,0.34901960784313724,0.8117647058823529}{{\textbf{define}}}}\ hello\\
{\textit{\ \ 2{\char58}\ }}\ \ (package\\
{\textit{\ \ 3{\char58}\ }}\ \ \ \ (name\ {\textcolor[rgb]{1.0,0.0,0.0}{"hello"}})\\
{\textit{\ \ 4{\char58}\ }}\ \ \ \ (version\ {\textcolor[rgb]{1.0,0.0,0.0}{"2.8"}})\\
{\textit{\ \ 5{\char58}\ }}\ \ \ \ (source\ (origin\ \ \ \ \ \ \ \ \ \ \ \ \ \ \ \ \ \ \ \ \ \ \ \ \ \ \ \ {\textcolor[rgb]{1.0,0.6509803921568628,0.0}{{\textbf{}}}}\\
{\textit{\ \ 6{\char58}\ }}\ \ \ \ \ \ \ \ \ \ \ \ \ \ (method\ url-fetch)\\
{\textit{\ \ 7{\char58}\ }}\ \ \ \ \ \ \ \ \ \ \ \ \ \ (uri\ (string-append\ {\textcolor[rgb]{1.0,0.0,0.0}{"mirror{\char58}//gnu/hello/hello-"}}\\
{\textit{\ \ 8{\char58}\ }}\ \ \ \ \ \ \ \ \ \ \ \ \ \ \ \ \ \ \ \ \ \ \ \ \ \ \ \ \ \ \ \ \ \ version\ {\textcolor[rgb]{1.0,0.0,0.0}{".tar.gz"}}))\ {\textcolor[rgb]{1.0,0.6509803921568628,0.0}{{\textbf{}}}}\\
{\textit{\ \ 9{\char58}\ }}\ \ \ \ \ \ \ \ \ \ \ \ \ \ (sha256\ (base32\ {\textcolor[rgb]{1.0,0.0,0.0}{"0wqd8..."}}))))\ \ \ {\textcolor[rgb]{1.0,0.6509803921568628,0.0}{{\textbf{}}}}\\
{\textit{\ 10{\char58}\ }}\ \ \ \ (build-system\ gnu-build-system)\ \ \ \ \ \ \ \ \ \ \ \ {\textcolor[rgb]{1.0,0.6509803921568628,0.0}{{\textbf{}}}}\\
{\textit{\ 11{\char58}\ }}\ \ \ \ (arguments\\
{\textit{\ 12{\char58}\ }}\ \ \ \ \ \ '(\#{\char58}configure-flags\ \ \ \ \ \ \ \ \ \ \ \ \ \ \ \ \ \ \ \ \ \ {\textcolor[rgb]{1.0,0.6509803921568628,0.0}{{\textbf{}}}}\\
{\textit{\ 13{\char58}\ }}\ \ \ \ \ \ \ \ `({\textcolor[rgb]{1.0,0.0,0.0}{"--disable-color"}}\\
{\textit{\ 14{\char58}\ }}\ \ \ \ \ \ \ \ \ \ ,(string-append\ {\textcolor[rgb]{1.0,0.0,0.0}{"--with-gawk="}}\\
{\textit{\ 15{\char58}\ }}\ \ \ \ \ \ \ \ \ \ \ \ \ \ \ \ \ \ \ \ \ \ \ \ \ \ (assoc-ref\ \%build-inputs\ {\textcolor[rgb]{1.0,0.0,0.0}{"gawk"}})))))\\
{\textit{\ 16{\char58}\ }}\ \ \ \ (inputs\ `(({\textcolor[rgb]{1.0,0.0,0.0}{"gawk"}}\ ,gawk)))\ \ \ \ \ \ \ \ \ \ \ \ \ \ \ \ {\textcolor[rgb]{1.0,0.6509803921568628,0.0}{{\textbf{}}}}\\
{\textit{\ 17{\char58}\ }}\ \ \ \ (synopsis\ {\textcolor[rgb]{1.0,0.0,0.0}{"GNU\ Hello"}})\\
{\textit{\ 18{\char58}\ }}\ \ \ \ (description\ {\textcolor[rgb]{1.0,0.0,0.0}{"An\ illustration\ of\ GNU's\ engineering\ practices."}})\\
{\textit{\ 19{\char58}\ }}\ \ \ \ (home-page\ {\textcolor[rgb]{1.0,0.0,0.0}{"http{\char58}//www.gnu.org/software/hello/"}})\\
{\textit{\ 20{\char58}\ }}\ \ \ \ (license\ gpl3+)))\\
\end{tabular}
}}
\egroup{\box1}}%
\setlength{\tabcolsep}{\oldpretabcolsep}
\caption{\label{fig:hello}A package definition using the high-level %
interface.}\end{figure*}
\noindent The interfaces described above remain fairly low-level.  In %
particular, they explicitly manipulate the store, pass around the system %
type, and are very distant from the abstract notion of a software %
package that we want to focus on.  To address this, Guix provides a %
high-level package definition interface.  It is designed to be {\em{purely declarative}} in common cases, while allowing users to customize %
the underlying build process.  That way, it should be intelligible and %
directly usable by packagers will little or no experience with Scheme. %
As an additional constraint, this extra layer should be efficient in %
space and time{\char58} package management tools need to be able to load and %
traverse a distribution consisting of thousands of packages.\par
\noindent Figure~\ref{fig:hello} shows the definition of %
the GNU~Hello package, a typical GNU package written in C and using %
the GNU build system---i.e., a {\texttt{configure}} script that generates a %
makefile supporting standardized targets such as {\texttt{check}} and {\texttt{install}}.  It is a direct mapping of the abstract notion of a software %
package and should be rather self-descriptive.\par
\noindent The {\texttt{inputs}} field %
specifies additional dependencies of the package.  Here line~16 means that Hello has a dependency labeled {\texttt{"gawk"}} on %
GNU~Awk, whose value is that of the {\texttt{gawk}} global variable; %
{\texttt{gawk}} is bound to a similar {\texttt{package}} declaration, omitted %
for conciseness.\par
\noindent The {\texttt{arguments}} field specifies arguments to be %
passed to the build system.  Here {\texttt{\#{\char58}configure-flags}}, %
unsurprisingly, specifies flags for the {\texttt{configure}} script.  Its %
value is quoted because it will be evaluated in the build %
stratum---i.e., in the build process, when the derivation is built.  It %
refers to the {\texttt{\%build-inputs}} global variable introduced in the %
build stratum by {\texttt{build-expression-$>$derivation}}, as seen before. %
That variable is bound to an association list that maps input names, %
like {\texttt{"gawk"}}, to their actual directory name on disk, like {\texttt{/\-nix/\-store/\-.\-.\-.\--gawk-4.\-0.\-2}}.\par
\noindent The code in Figure~\ref{fig:hello} demonstrates %
Guix's use of embedded domain-specific languages (EDSLs).  The {\texttt{package}} form, the {\texttt{origin}} form (line 5), %
and the {\texttt{base32}} form (line 9) are expanded at %
macro-expansion time.  The {\texttt{package}} and {\texttt{origin}} forms %
expand to a call to Guile's {\texttt{make-struct}} primitive, which instantiates a record of the given type %
and with the given field values\footnote{The {\texttt{make-struct}} %
instantiates %
SRFI-9-style flat records, which are essentially vectors of a disjoint %
type.  In Guile they are lightweight compared to CLOS-style objects, %
both in terms of run time and memory footprint.  Furthermore, {\texttt{make-struct}} is subject to inlining.}; these macros look up %
the mapping of field names to field indexes, such that that mapping %
incurs no run-time overhead, in a way similar to SRFI-35 records [14].  They also bind fields as per {\texttt{letrec*}}, %
allowing them to refer to one another, as on line 8 %
of Figure~\ref{fig:hello}.  The {\texttt{base32}} macro simply converts a literal %
string containing a base-32 representation into a bytevector literal, %
again allowing the conversion and error-checking to be done at expansion %
time rather than at run-time.\par
\begin{figure}[ht]
\setlength{\oldpretabcolsep}{\tabcolsep}
\addtolength{\tabcolsep}{-\tabcolsep}
{\setbox1 \vbox \bgroup
{\noindent \texttt{\begin{tabular}{l}
{\textit{\ \ 1{\char58}\ }}(define-record-type*\ $<$package$>$\\
{\textit{\ \ 2{\char58}\ }}\ \ package\ make-package\\
{\textit{\ \ 3{\char58}\ }}\ \ package?\\
{\textit{\ \ 4{\char58}\ }}\\
{\textit{\ \ 5{\char58}\ }}\ \ (name\ package-name)\\
{\textit{\ \ 6{\char58}\ }}\ \ (version\ package-version)\\
{\textit{\ \ 7{\char58}\ }}\ \ (source\ package-source)\\
{\textit{\ \ 8{\char58}\ }}\ \ (build-system\ package-build-system)\\
{\textit{\ \ 9{\char58}\ }}\ \ (arguments\ package-arguments\\
{\textit{\ 10{\char58}\ }}\ \ \ \ \ \ \ \ \ \ \ \ \ (default\ '())\ (thunked))\\
{\textit{\ 11{\char58}\ }}\\
{\textit{\ 12{\char58}\ }}\ \ (inputs\ package-inputs\\
{\textit{\ 13{\char58}\ }}\ \ \ \ \ \ \ \ \ \ (default\ '())\ (thunked))\ \ \ \ \ \ \ \ \ \ \ \ \ {\textcolor[rgb]{1.0,0.6509803921568628,0.0}{{\textbf{}}}}\\
{\textit{\ 14{\char58}\ }}\ \ (propagated-inputs\ package-propagated-inputs\\
{\textit{\ 15{\char58}\ }}\ \ \ \ \ \ \ \ \ \ \ \ \ \ \ \ \ \ \ \ \ (default\ '()))\\
{\textit{\ 16{\char58}\ }}\\
{\textit{\ 17{\char58}\ }}\ \ (synopsis\ package-synopsis)\\
{\textit{\ 18{\char58}\ }}\ \ (description\ package-description)\\
{\textit{\ 19{\char58}\ }}\ \ (license\ package-license)\\
{\textit{\ 20{\char58}\ }}\ \ (home-page\ package-home-page)\\
{\textit{\ 21{\char58}\ }}\\
{\textit{\ 22{\char58}\ }}\ \ (location\ package-location\\
{\textit{\ 23{\char58}\ }}\ \ \ \ \ (default\ (current-source-location))))\\
\end{tabular}
}}
\egroup{\box1}}%
\setlength{\tabcolsep}{\oldpretabcolsep}
\caption{\label{fig:package}Definition of the {\texttt{package}} record type.}\end{figure}
\noindent The {\texttt{package}} and {\texttt{origin}} macros are generated %
by a {\texttt{syntax-case}} hygienic macro [19], %
{\texttt{define-record-type*}}, which is layered above SRFI-9's syntactic %
record layer [13].  Figure~\ref{fig:package} shows the definition of the {\texttt{$<$package$>$}} record %
type (the {\texttt{$<$origin$>$}} record type, not shown here, is defined %
similarly.)  In addition to the name of a procedural constructor, {\texttt{make-package}}, as with SRFI-9, the name of a {\em{syntactic}} %
constructor, {\texttt{package}}, is given (likewise, {\texttt{origin}} is the %
syntactic constructor of {\texttt{$<$origin$>$}}.)  Fields may have a default %
value, introduced with the {\texttt{default}} keyword.  An interesting use %
of default values is the {\texttt{location}} field{\char58} its default value is %
the result of {\texttt{current-source-location}}, which is itself a %
built-in macro that expands to the source file location of the {\texttt{package}} form.  Thus, records defined with the {\texttt{package}} macro %
automatically have a {\texttt{location}} field denoting their source file %
location.  This allows the user interface to report source file location %
in error messages and in package search results, thereby making it %
easier for users to ``jump into'' the distribution's source, which is %
one of our goals.\par
\begin{figure}[ht]
\setlength{\oldpretabcolsep}{\tabcolsep}
\addtolength{\tabcolsep}{-\tabcolsep}
{\setbox1 \vbox \bgroup
{\noindent \texttt{\begin{tabular}{l}
{\textit{\ \ 1{\char58}\ }}(package\ (inherit\ hello)\\
{\textit{\ \ 2{\char58}\ }}\ \ (version\ {\textcolor[rgb]{1.0,0.0,0.0}{"2.7"}})\\
{\textit{\ \ 3{\char58}\ }}\ \ (source\\
{\textit{\ \ 4{\char58}\ }}\ \ \ \ (origin\\
{\textit{\ \ 5{\char58}\ }}\ \ \ \ \ \ (method\ url-fetch)\\
{\textit{\ \ 6{\char58}\ }}\ \ \ \ \ \ (uri\\
{\textit{\ \ 7{\char58}\ }}\ \ \ \ \ \ \ {\textcolor[rgb]{1.0,0.0,0.0}{"mirror{\char58}//gnu/hello/hello-2.7.tar.gz"}})\\
{\textit{\ \ 8{\char58}\ }}\ \ \ \ \ \ (sha256\\
{\textit{\ \ 9{\char58}\ }}\ \ \ \ \ \ \ \ (base32\ {\textcolor[rgb]{1.0,0.0,0.0}{"7dqw3..."}})))))\\
\end{tabular}
}}
\egroup{\box1}}%
\setlength{\tabcolsep}{\oldpretabcolsep}
\caption{\label{fig:inherit}Creating a variant of the {\texttt{hello}} package.}\end{figure}
\noindent The syntactic constructors generated by {\texttt{define-record-type*}} additionally support a form of {\em{functional %
setters}} (sometimes referred to as ``lenses'' [15]), {\textit{via}} the {\texttt{inherit}} keyword.  It allows %
programmers to create new instances that differ from an existing %
instance by one or more field values.  A typical use case is shown %
in Figure~\ref{fig:inherit}{\char58} the expression shown evaluates %
to a new {\texttt{$<$package$>$}} instance whose fields all have the same value %
as the {\texttt{hello}} variable of Figure~\ref{fig:hello}, %
except for the {\texttt{version}} and {\texttt{source}} fields.  Under the %
hood, again, this expands to a single {\texttt{make-struct}} call with {\texttt{struct-ref}} calls for fields whose value is reused.\par
\noindent The {\texttt{inherit}} feature supports a very useful idiom.  It %
allows new package variants to be created programmatically, concisely, %
and in a purely functional way.  It is notably used to bootstrap the %
software distribution, where bootstrap variants of packages such as %
GCC or the GNU~libc are %
built with different inputs and configuration flags than the %
final versions.  Users can similarly define customized variants of the %
packages found in the distribution. %
This feature also allows high-level transformations to be %
implemented as pure functions.  For instance, the {\texttt{static-package}} %
procedure takes a {\texttt{$<$package$>$}} instance, and returns a variant of %
that package that is statically linked.  It operates by just adding the %
relevant {\texttt{configure}} flags, and recursively applying itself to the %
package's inputs.\par
\noindent Another application is the {\em{on-line auto-updater}}{\char58} when %
installing a GNU package defined in the distribution, the {\texttt{guix %
package}} command automatically %
checks whether a newer version is available upstream from {\texttt{ftp.\-gnu.\-org}}, and offers the option to substitute the package's source %
with a fresh download of the %
new upstream version---all at run time.This kind of feature is hardly accessible to an external DSL %
implementation.  Among other things, this feature requires networking %
primitives (for the FTP client), which are typically unavailable in an %
external DSL such as the Nix language.  The feature could be implemented %
in a language other than the DSL---for instance, Nix can export its %
abstract syntax tree as XML to external programs.  However, this %
approach is often inefficient, due to the format conversion, and lossy{\char58} %
the exported representation may be either be too distant from the source %
code, or too distant from the preferred abstraction level.  The author's %
own experience writing an off-line auto-updater for Nix revealed other %
specific issues; for instance, the Nix language is lazily evaluated, but %
to make use of its XML output, one has to force strict evaluation, which %
in turn may generate more data than needed.  In Guix, {\texttt{$<$package$>$}} %
instances have the expected level of abstraction, and they are readily %
accessible as first-class Scheme objects.\par
\noindent Sometimes it is desirable for the value of a field to depend %
on the system type targeted.  For instance, for bootstrapping purposes, %
MIT/GNU~Scheme's build system depends on pre-compiled binaries, which %
are architecture-dependent; its {\texttt{input}} field must be able to %
select the right binaries depending on the architecture. %
To allow field values to refer to the target system type, %
we resort to {\em{thunked}} fields, as shown on line 13 of Figure~\ref{fig:package}.  These fields have %
their value automatically wrapped in a thunk (a zero-argument %
procedure); when accessing them with the associated accessor, the thunk %
is transparently invoked.  Thus, the values of thunked fields are %
computed lazily; more to the point, they can refer to {\em{dynamic %
state}} in place at their invocation point.  In particular, the {\texttt{package-derivation}} procedure (shortly introduced) sets up a {\texttt{current-system}} dynamically-scoped parameter, which allows field %
values to know what the target system is.\par
\noindent Finally, both {\texttt{$<$package$>$}} and {\texttt{$<$origin$>$}} records %
have an associated ``compiler'' that turns them into a derivation.  {\texttt{origin-derivation}} takes an {\texttt{$<$origin$>$}} instance and returns a %
derivation that downloads it, according to its {\texttt{method}} field. %
Likewise, {\texttt{package-derivation}} takes a package and returns a %
derivation that builds it, according to its {\texttt{build-system}} and %
associated {\texttt{arguments}} (more on that \hyperref[shell]{in Section~3.4}).  As we have seen on Figure~\ref{fig:hello}, the {\texttt{inputs}} field lists dependencies of a package, %
which are themselves {\texttt{$<$package$>$}} objects; the {\texttt{package-derivation}} procedure recursively applies to those inputs, %
such that their derivation is computed and passed as the inputs argument %
of the lower-level {\texttt{build-expression-$>$derivation}}.\par
\noindent Guix essentially implements {\em{deep embedding}} of DSLs, %
where the semantics of the packaging DSL is interpreted by a dedicated %
compiler [12].  Of course the DSLs defined here %
are simple, but they illustrate how Scheme's primitive mechanisms, %
in particular macros, make it easy to implement such DSLs without %
requiring any special support from the Scheme implementation.\par

\subsection{Build Programs}
\label{shell}
\noindent The value of the {\texttt{build-system}} field, as shown on %
Figure~\ref{fig:hello}, must be a {\texttt{build-system}} %
object, which is essentially a wrapper around two procedure{\char58} one %
procedure to do a native build, and one to do a cross-build.  When the %
aforementioned {\texttt{package-derivation}} (or {\texttt{package-cross-derivation}}, when cross-building) is called, it invokes %
the build system's build procedure, passing it a connection to the %
build daemon, the system type, derivation name, and inputs.  It is the %
build system's responsibility to return a derivation that actually %
builds the software.\par
\begin{figure}[ht]
\setlength{\oldpretabcolsep}{\tabcolsep}
\addtolength{\tabcolsep}{-\tabcolsep}
{\setbox1 \vbox \bgroup
{\noindent \texttt{\begin{tabular}{l}
(define*\ (gnu-build\ \#{\char58}key\ (phases\ \%standard-phases)\\
\ \ \ \ \ \ \ \ \ \ \ \ \ \ \ \ \ \ \ \ \#{\char58}allow-other-keys\\
\ \ \ \ \ \ \ \ \ \ \ \ \ \ \ \ \ \ \ \ \#{\char58}rest\ args)\\
\ \ ;;\ Run\ all\ the\ PHASES\ in\ order,\ passing\ them\ ARGS.\\
\ \ ;;\ Return\ true\ on\ success.\\
\ \ (every\ (match-lambda\\
\ \ \ \ \ \ \ \ \ \ ((name\ .\ proc)\\
\ \ \ \ \ \ \ \ \ \ \ (format\ \#t\ "starting\ phase\ `$_{\mbox{\char126}}$a'$_{\mbox{\char126}}$\%"\ name)\\
\ \ \ \ \ \ \ \ \ \ \ (let\ ((result\ (apply\ proc\ args)))\\
\ \ \ \ \ \ \ \ \ \ \ \ \ (format\ \#t\ "phase\ `$_{\mbox{\char126}}$a'\ done$_{\mbox{\char126}}$\%"\ name)\\
\ \ \ \ \ \ \ \ \ \ \ \ \ result)))\\
\ \ \ \ \ \ \ \ \ phases))\\
\end{tabular}
}}
\egroup{\box1}}%
\setlength{\tabcolsep}{\oldpretabcolsep}
\caption{\label{fig:gnu-build}Entry point of the builder side code of {\texttt{gnu-build-system}}.}\end{figure}
\noindent The {\texttt{gnu-build-system}} object (line 10 of Figure~\ref{fig:hello}) provides %
procedures to build and cross-build software that uses the GNU build %
system or similar.  In a nutshell, it runs the following phases by %
default{\char58} %
\begin{enumerate}
 \item unpack the source tarball, and change the current directory %
to the resulting directory;
 \item patch shebangs on installed files---e.g., replace {\texttt{\#!/\-bin/\-sh}} by {\texttt{\#!/\-nix/\-store/\-.\-.\-.\--bash-4.\-2/\-bin/\-sh}}; this is %
required to allow scripts to work with our unusual file system layout;
 \item run {\texttt{.\-/\-configure {-}{-}prefix=/\-nix/\-store/\-.\-.\-.\-}}, followed by %
{\texttt{make}} and {\texttt{make check}}
 \item run {\texttt{make install}} and patch shebangs in installed files.
\end{enumerate}
Of course, that is all implemented in Scheme, {\textit{via}} {\texttt{build-expression-$>$derivation}}.  Supporting code is available as a %
build-side module that {\texttt{gnu-build-system}} automatically adds as an %
input to its build scripts.  The default build programs just call the %
procedure of that module that runs the above phases.\par
\noindent The {\texttt{(guix build gnu-build-system)}} module contains %
the implementation of the above phases; it is imported on the builder %
side.  The phases are modeled as follows{\char58} each phase is a procedure %
accepting several keyword arguments, and ignoring any keyword arguments %
it does not recognize\footnote{Like many Scheme implementations, Guile %
supports {\em{named}} or {\em{keyword}} arguments as an extension %
to the R5 and R6RS.  In addition, procedure definitions whose formal %
argument list contains the {\texttt{\#{\char58}allow-other-keys}} keyword ignore any %
unrecognized keyword arguments that they are passed.}.  For instance %
the {\texttt{configure}} procedure is in charge of running the package's %
{\texttt{.\-/\-configure}} script; that procedure honors the {\texttt{\#{\char58}configure-flags}} keyword parameter seen on Figure~\ref{fig:hello}.  Similarly, the {\texttt{build}}, {\texttt{check}}, and {\texttt{install}} procedures run the {\texttt{make}} command, and all honor the %
{\texttt{\#{\char58}make-flags}} keyword parameter.\par
\noindent All the procedures implementing the standard phases of the %
GNU build system are listed in the {\texttt{\%standard-phases}} builder-side %
variable, in the form of a list of phase name\-/procedure pairs.  The %
entry point of the builder-side code of {\texttt{gnu-build-system}} is %
shown on Figure~\ref{fig:gnu-build}.  It calls all the %
phase procedures in order, by default those listed in the {\texttt{\%standard-phases}} association list, passing them all the arguments it %
got; its return value is true when every procedure's return value is %
true.\par
\begin{figure}[ht]
\setlength{\oldpretabcolsep}{\tabcolsep}
\addtolength{\tabcolsep}{-\tabcolsep}
{\setbox1 \vbox \bgroup
{\noindent \texttt{\begin{tabular}{l}
(define\ howdy\\
\ \ (package\ (inherit\ hello)\\
\ \ \ \ (arguments\\
\ \ \ \ \ \ '(\#{\char58}phases\\
\ \ \ \ \ \ \ \ (alist-cons-after\\
\ \ \ \ \ \ \ \ \ \ 'configure\ 'change-hello\\
\ \ \ \ \ \ \ \ \ \ (lambda*\ (\#{\char58}key\ system\ \#{\char58}allow-other-keys)\\
\ \ \ \ \ \ \ \ \ \ \ \ (substitute*\ "src/hello.c"\\
\ \ \ \ \ \ \ \ \ \ \ \ \ \ (("Hello,\ world!")\\
\ \ \ \ \ \ \ \ \ \ \ \ \ \ \ (string-append\ "Howdy!\ Running\ on\ "\\
\ \ \ \ \ \ \ \ \ \ \ \ \ \ \ \ \ \ \ \ \ \ \ \ \ \ \ \ \ \ system\ "."))))\\
\ \ \ \ \ \ \ \ \ \ \%standard-phases)))))\\
\end{tabular}
}}
\egroup{\box1}}%
\setlength{\tabcolsep}{\oldpretabcolsep}
\caption{\label{fig:hello-custom}Package specification with custom build phases.}\end{figure}
\noindent The {\texttt{arguments}} field, shown on Figure~\ref{fig:hello}, allows users to pass keyword arguments to the %
builder-side code.  In addition to the {\texttt{\#{\char58}configure-flags}} %
argument shown on the figure, users may use the {\texttt{\#{\char58}phases}} %
argument to specify a different set of phases.  The value of the {\texttt{\#{\char58}phases}} must be a list of phase name/procedure pairs, as discussed %
above.  This allows users to arbitrarily extend or modify the behavior %
of the build system.  Figure~\ref{fig:hello-custom} shows %
a variant of the definition in Figure~\ref{fig:hello} that %
adds a custom build phase.  The {\texttt{alist-cons-after}} procedure is %
used to add a pair with {\texttt{change-hello}} as its first item and the %
{\texttt{lambda*}} as its second item right after the pair in {\texttt{\%standard-phases}} whose first item is {\texttt{configure}}; in other %
words, it reuses the standard build phases, but with an additional {\texttt{change-hello}} phase right after the {\texttt{configure}} phase.  The %
whole {\texttt{alist-cons-after}} expression is evaluated on the %
builder side.\par
\noindent This approach was inspired by that of NixOS, which uses %
Bash for its build scripts.  Even with ``advanced'' Bash features such %
as functions, arrays, and associative arrays, the phases mechanism in %
NixOS remains limited and fragile, often leading to string escaping %
issues and obscure error reports due to the use of {\texttt{eval}}.  Again, %
using Scheme instead of Bash unsurprisingly allows for better code %
structuring, and improves flexibility.\par
\noindent Other build systems are provided.  For instance, the standard %
build procedure for Perl packages is slightly different{\char58} mainly, the %
configuration phase consists in running {\texttt{perl Makefile.\-PL}}, and %
test suites are run with {\texttt{make test}} instead of {\texttt{make %
check}}.  To accommodate that, Guix provides {\texttt{perl-build-system}}. %
Its companion build-side module essentially calls out to that of {\texttt{gnu-build-system}}, only with appropriate {\texttt{configure}} and {\texttt{check}} phases.  This mechanism is similarly used %
for other build systems such as CMake and Python's build system.\par
\begin{figure}[ht]
\setlength{\oldpretabcolsep}{\tabcolsep}
\addtolength{\tabcolsep}{-\tabcolsep}
{\setbox1 \vbox \bgroup
{\noindent \texttt{\begin{tabular}{l}
(substitute*\ (find-files\ "gcc/config"\\
\ \ \ \ \ \ \ \ \ \ \ \ \ \ \ \ \ \ \ \ \ \ \ \ \ "{\char94}gnu-user(64)?{\char92}{\char92}.h\$")\\
\ \ (("\#define\ LIB\_SPEC\ (.*)\$"\ \_\ suffix)\\
\ \ \ (string-append\ "\#define\ LIB\_SPEC\ {\char92}"-L"\ libc\\
\ \ \ \ \ \ \ \ \ \ \ \ \ \ \ \ \ \ "/lib\ {\char92}"\ "\ suffix\ "{\char92}n"))\\
\ \ (("\#define\ STARTFILE\_SPEC.*\$"\ line)\\
\ \ \ (string-append\ "\#define\ STARTFILE\_PREFIX\_1\ {\char92}""\\
\ \ \ \ \ \ \ \ \ \ \ \ \ \ \ \ \ \ libc\ "/lib{\char92}"{\char92}n"\ line)))\\
\end{tabular}
}}
\egroup{\box1}}%
\setlength{\tabcolsep}{\oldpretabcolsep}
\caption{\label{fig:substitute*}The {\texttt{substitute*}} macro for {\texttt{sed}}-like substitutions.}\end{figure}
\noindent Build programs often need to traverse file trees, modify %
files according to a given pattern, etc.  One example is the ``patch %
shebang'' phase mentioned above{\char58} all the source files must be traversed, %
and those starting with {\texttt{\#!}} are candidate to patching.  This kind %
of task is usually associated with ``shell programming''---as is the %
case with the build scripts found in NixOS, which are written in Bash, %
and resort to {\texttt{sed}}, {\texttt{find}}, etc.  In Guix, a build-side %
Scheme module provides the necessary tools, built on top of Guile's %
operating system interface.  For instance, {\texttt{find-files}} returns a %
list of files whose names matches a given pattern; {\texttt{patch-shebang}} %
performs the {\texttt{\#!}} adjustment described above; {\texttt{copy-recursively}} and {\texttt{delete-recursively}} are the equivalent, %
respectively, of the shell {\texttt{cp -r}} and {\texttt{rm -rf}} commands; %
etc.\par
\noindent An interesting example is the {\texttt{substitute*}} macro, %
which does {\texttt{sed}}-style substitution on files.  Figure~\ref{fig:substitute*} illustrates its use to patch a series of %
files returned by {\texttt{find-files}}.  There are two clauses, each with %
a pattern in the form of a POSIX regular expression; each clause's body %
returns a string, which is the substitution for any matching line in the %
given files.  In the first clause's body, {\texttt{suffix}} is bound to the %
submatch corresponding to {\texttt{(.\-*)}} in the regexp; in the second %
clause, {\texttt{line}} is bound to the whole match for that regexp.  This %
snippet is nearly as concise than equivalent shell code using {\texttt{find}} and {\texttt{sed}}, and it is much easier to work with.\par
\noindent Build-side modules also include support for fetching files %
over HTTP (using Guile's web client module) and FTP, as needed to %
realize the derivation of {\texttt{origin}}s (line 5 of %
Figure~\ref{fig:hello}).  TLS support is available when %
needed through the Guile bindings of the GnuTLS library.\par

\section{Evaluation and Discussion}
\label{eval}
\noindent This section discusses the current status of Guix and its %
associated GNU/Linux distribution, and outlines key aspects of their %
development.\par

\subsection{Status}
\label{section8851}
\noindent Guix is still a young project.  Its main features as a %
package manager are already available.  This includes the APIs discussed %
in \hyperref[api]{Section~3}, as well as command-line %
interfaces.  The development of Guix's interfaces was facilitated by the %
reuse of Nix's build daemon as the storage and deployment layer.\par
\noindent The {\texttt{guix package}} command is the main user %
interface{\char58} it allows packages to be browsed, installed, removed, and %
upgraded.  The command takes care of maintaining meta-data about %
installed packages, as well as a per-user tree of symlinks pointing to %
the actual package files in {\texttt{/\-nix/\-store}}, called the {\em{user %
profile}}.  It has a simple interface.  For instance, the following %
command installs Guile and removes Bigloo from the user's profile, as a %
single transaction{\char58}
\begin{quote}
\setlength{\oldpretabcolsep}{\tabcolsep}
\addtolength{\tabcolsep}{-\tabcolsep}
{\setbox1 \vbox \bgroup
{\noindent \texttt{\begin{tabular}{l}
\$\ guix\ package\ {-}{-}install\ guile\ {-}{-}remove\ bigloo\\
\end{tabular}
}}
\egroup{\box1}}%
\setlength{\tabcolsep}{\oldpretabcolsep}

\end{quote}The transaction can be rolled back with the following command{\char58}
\begin{quote}
\setlength{\oldpretabcolsep}{\tabcolsep}
\addtolength{\tabcolsep}{-\tabcolsep}
{\setbox1 \vbox \bgroup
{\noindent \texttt{\begin{tabular}{l}
\$\ guix\ package\ {-}{-}roll-back\\
\end{tabular}
}}
\egroup{\box1}}%
\setlength{\tabcolsep}{\oldpretabcolsep}

\end{quote}The following command upgrades all the installed packages %
whose name starts with a `g'{\char58}
\begin{quote}
\setlength{\oldpretabcolsep}{\tabcolsep}
\addtolength{\tabcolsep}{-\tabcolsep}
{\setbox1 \vbox \bgroup
{\noindent \texttt{\begin{tabular}{l}
\$\ guix\ package\ {-}{-}upgrade\ '{\char94}g.*'\\
\end{tabular}
}}
\egroup{\box1}}%
\setlength{\tabcolsep}{\oldpretabcolsep}

\end{quote}The {\texttt{{-}{-}list-installed}} and {\texttt{{-}{-}list-available}} %
options can be used to list the installed or available packages.\par
\noindent As of this writing, Guix comes with a user-land %
distribution of GNU/Linux.  That is, it allows users to install packages %
on top of a running GNU/Linux system.  The distribution is %
self-contained, as explained in \hyperref[bootstrap]{Section~4.3}, and available on {\texttt{x86\_64}} and {\texttt{i686}}.  It provides more than 400 %
packages, including core GNU packages such as the GNU C Library, GCC, %
Binutils, and Coreutils, as well as the Xorg software stack and %
applications such as Emacs, TeX~Live, and several Scheme %
implementations.  This is roughly a tenth of the number of packages %
found in mature free software distributions such as Debian.  Experience %
with NixOS suggests that the functional model, coupled with continuous %
integration, allows the distribution to grow relatively quickly, because it is %
always possible to precisely monitor the status of the whole %
distribution and the effect of a change---unlike with imperative %
distributions, where the upgrade of a single package can affect many %
applications in many unpredictable ways~[7].\par
\noindent From a programming point of view, packages are exposed as %
first-class global variables.  For instance, the {\texttt{(gnu packages %
guile)}} module exports two variables, {\texttt{guile-1.\-8}} and {\texttt{guile-2.\-0}}, each bound to a {\texttt{$<$package$>$}} variable corresponding %
to the legacy and current stable series of Guile.  In turn, this module %
imports {\texttt{(gnu packages multiprecision)}}, which exports a {\texttt{gmp}} global variable, among other things; that {\texttt{gmp}} variable is %
listed in the {\texttt{inputs}} field of {\texttt{guile}} and {\texttt{guile-2.\-0}}.  The package manager {\em{and}} the distribution are %
just a set of ``normal'' modules that any program or library can use.\par
\noindent Packages carry meta-data, as shown in Figure~\ref{fig:hello}.  Synopses and descriptions are internationalized using %
GNU~Gettext---that is, they can be translated in the user's native %
language, a feature that comes for free when embedding the DSL in a %
mature environment like Guile.  We are in the process of implementing %
mechanisms to synchronize part of that meta-data, such as synopses, with %
other databases of the GNU Project.\par
\noindent While the distribution is not bootable yet, it already %
includes a set of tools to build bootable GNU/Linux images for the QEMU %
emulator.  This includes a package for the kernel itself, as well as %
procedures to build QEMU images, and Linux ``initrd''---the ``initial %
RAM disk'' used by Linux when booting, and which is responsible for %
loading essential kernel modules and mounting the root file system, %
among other things.  For example, we provide the {\texttt{expression-$>$derivation-in-linux-vm}}{\char58} it works in a way similar to {\texttt{build-expression-$>$derivation}}, except that the given expression is %
evaluated in a virtual machine that mounts the host's store over CIFS. %
As a demonstration, we implemented a derivation that builds a %
``boot-to-Guile'' QEMU image, where the initrd contains a %
statically-linked Guile that directly runs a boot program written in %
Scheme [5].\par
\noindent The performance-critical parts are the derivation %
primitives discussed in \hyperref[api]{Section~3}.  For %
instance, the computation of Emacs's derivation involves that of 292 %
other derivations---that is, 292 invocations of the {\texttt{derivation}} primitive---co\-rres\-pon\-ding to 582 RPCs\footnote{The %
number of {\texttt{derivation}} calls and {\texttt{add-to-store}} RPCs is %
reduced thanks to the use of client-side memoization.}.  The wall time %
of evaluating that derivation is 1.1 second on average on a 2.6~GHz %
{\texttt{x86\_64}} machine.  This is acceptable as a user, but 5 times %
slower than Nix's clients for a similar derivation written in the Nix %
language.  Profiling shows that %
Guix spends most of its time in its derivation serialization code and %
RPCs.  We interpret this as a consequence of Guix's unoptimized code, as %
well as the difference between native C++ code and our interpreted %
bytecode.\par

\subsection{Purity}
\label{section8878}
\noindent Providing pure build environments that do not honor the %
``standard'' file system layout turned out not to be a problem, as %
already evidenced in NixOS [8].  This is largely %
thanks to the ubiquity of the GNU build system, which strives to %
provide users with ways to customize the layout of installed %
packages and to adjust to the user's file locations.\par
\noindent The only directories visible in the build {\texttt{chroot}} %
environment are {\texttt{/\-dev}}, {\texttt{/\-proc}}, and the subset of {\texttt{/\-nix/\-store}} that is explicitly declared in the derivation being built. %
NixOS makes one exception{\char58} it relies on the availability of {\texttt{/\-bin/\-sh}} in the {\texttt{chroot}} [9]. %
We remove that exception, and instead %
automatically patch script ``shebangs'' in the package's source, as %
noted in \hyperref[shell]{Section~3.4}.  This turned out to %
be more than just a theoretical quest for ``purity''.  First, some %
GNU/Linux distributions use Dash as the implementation of {\texttt{/\-bin/\-sh}}, while %
others use Bash; these are two variants of the Bourne shell, with %
different extensions, and in general different behavior.  Second, {\texttt{/\-bin/\-sh}} is typically a dynamically-linked executable.  So adding {\texttt{/\-bin}} to the {\texttt{chroot}} is not enough; one typically needs to also %
add {\texttt{/\-lib*}} and {\texttt{/\-lib/\-*-linux-gnu}} to the chroot.  At that %
point, there are many impurities, and a great potential for %
non-reproducibility---which defeats the purpose of the {\texttt{chroot}}.\par
\noindent Several packages had to be adjusted for proper function in %
the absence of {\texttt{/\-bin/\-sh}} [6].  In %
particular, libc's {\texttt{system}} and {\texttt{popen}} functions had to be %
changed to refer to ``our'' Bash instance.  Likewise, GNU~Make, %
GNU~Awk, GNU~Guile, and Python needed adjustment.  Occasionally, %
occurrences of {\texttt{/\-bin/\-sh}} are not be handled automatically, for %
instance in test suites; these have to be patched manually in the %
package's recipe.\par

\subsection{Bootstrapping}
\label{bootstrap}
\begin{figure*}[!th]

\includegraphics[width=1.0\textwidth]{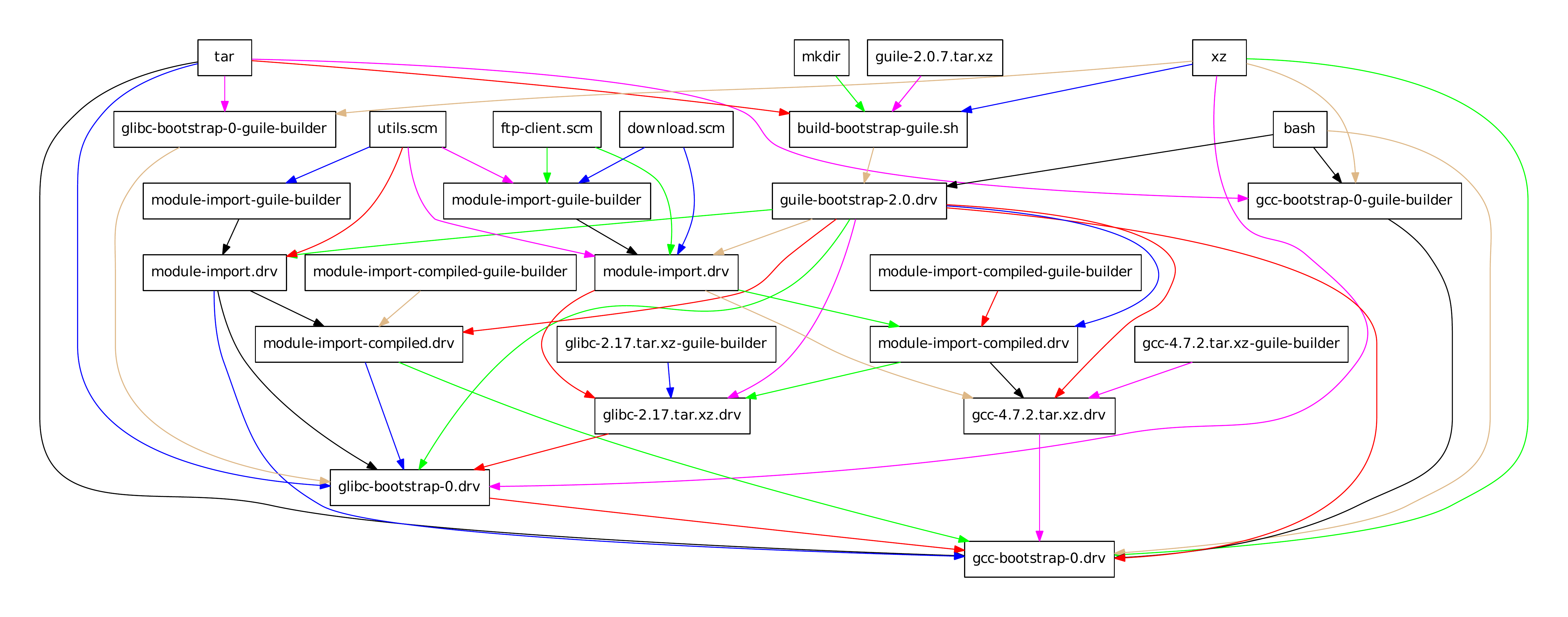}
\caption{\label{boot-graph}Dependency graph of the software distribution %
bootstrap.}\end{figure*}
\noindent Bootstrapping in our context refers to how the distribution %
gets built ``from nothing''.  Remember that the build environment of a %
derivation contains nothing but its declared inputs.  So there's an %
obvious chicken-and-egg problem{\char58} how does the first package get built? %
How does the first compiler get compiled?\par
\noindent The GNU system we are building is primarily made of C code, %
with libc at its core.  The GNU build system itself assumes the %
availability of a Bourne shell, traditional Unix tools provided by %
GNU~Coreutils, Awk, Findutils, sed, and grep.  Furthermore, our build %
programs are written in Guile Scheme.  Consequently, we rely on %
pre-built statically-linked binaries of GCC, Binutils, libc, and the %
other packages mentioned above to get started.\par
\noindent Figure~\ref{boot-graph} shows the very beginning %
of the dependency graph of our distribution.  At this level of detail, %
things are slightly more complex.  First, Guile itself consists of an %
ELF executable, along with many source and compiled Scheme files that %
are dynamically loaded when it runs.  This gets stored in the {\texttt{guile-2.\-0.\-7.\-tar.\-xz}} tarball shown in this graph.  This tarball is part %
of Guix's ``source'' distribution, and gets inserted into the store with %
{\texttt{add-to-store}}.\par
\noindent But how do we write a derivation that unpacks this tarball %
and adds it to the store?  To solve this problem, the {\texttt{guile-bootstrap-2.\-0.\-drv}} derivation---the first one that gets %
built---uses {\texttt{bash}} as its builder, which runs {\texttt{build-bootstrap-guile.\-sh}}, which in turn calls {\texttt{tar}} to unpack %
the tarball.  Thus, {\texttt{bash}}, {\texttt{tar}}, {\texttt{xz}}, and {\texttt{mkdir}} are statically-linked binaries, also part of the Guix source %
distribution, whose sole purpose is to allow the Guile tarball to be %
unpacked.\par
\noindent Once {\texttt{guile-bootstrap-2.\-0.\-drv}} is built, we have a %
functioning Guile that can be used to run subsequent build programs. %
Its first task is to download tarballs containing the other pre-built %
binaries---this is what the {\texttt{.\-tar.\-xz.\-drv}} derivations do. %
Guix modules such as {\texttt{ftp-client.\-scm}} are used for this purpose. %
The {\texttt{module-import.\-drv}} derivations %
import those modules in a directory in the store, using the original %
layout\footnote{In Guile, module names are a list of symbols, such as %
{\texttt{(guix ftp-client)}}, which map directly to file names, such as %
{\texttt{guix/\-ftp-client.\-scm}}.}.  The {\texttt{module-import-compiled.\-drv}} %
derivations compile those modules, and write them in an output directory %
with the right layout.  This corresponds to the {\texttt{\#{\char58}module}} %
argument of {\texttt{build-expression-$>$derivation}} mentioned in \hyperref[build-exprs]{Section~3.2}.\par
\noindent Finally, the various tarballs are unpacked by the %
derivations {\texttt{gcc-bootstrap-0.\-drv}}, {\texttt{glibc-bootstrap-0.\-drv}}, %
etc., at which point we have a working C GNU tool chain.  The first %
tool that gets built with these tools (not shown here) is GNU~Make, %
which is a prerequisite for all the following packages.\par
\noindent Bootstrapping is complete when we have a full tool %
chain that does not depend on the pre-built bootstrap tools shown in %
Figure~\ref{boot-graph}.  Ways to achieve this are known, and %
notably documented by the {\textit{Linux From Scratch}} project [1].  We can formally verify this no-dependency %
requirement by checking whether the files of the final tool chain %
contain references to the {\texttt{/\-nix/\-store}} directories of the %
bootstrap inputs.\par
\noindent Obviously, Guix contains {\texttt{package}} declarations to %
build the bootstrap binaries shown in Figure~\ref{boot-graph}. %
Because the final tool chain does not depend on those tools, they rarely %
need to be updated.  Having a way to do that automatically proves to be %
useful, though.  Coupled with Guix's nascent support for %
cross-compilation, porting to a new architecture will boil down to %
cross-building all these bootstrap tools.\par

\section{Related Work}
\label{related}
\noindent Numerous package managers for Scheme programs and libraries %
have been developed, including Racket's PLaneT, Dorodango for R6RS %
implementations, Chicken Scheme's ``Eggs'', Guildhall for Guile, and %
ScmPkg [16].  Unlike GNU~Guix, they are %
typically limited to Scheme-only code, and take the core operating %
system software for granted.  To our knowledge, they implement the %
{\em{imperative}} package management paradigm, and do not attempt to %
support features such as transactional upgrades and rollbacks. %
Unsurprisingly, these tools rely on package descriptions that more or %
less resemble those described in \hyperref[pkg-decl]{Section~3.3}; however, in the case of at least ScmPkg, Dorodango, and %
Guildhall, package descriptions are written in an {\em{external}} %
DSL, which happens to use s-expression syntax.\par
\noindent In [21], the authors illustrate how the %
{\textit{units}} mechanism of MzScheme modules could be leveraged to %
improve operating system packaging systems.  The examples therein focus %
on OS services, and multiple instantiation thereof, rather than on %
package builds and composition.\par
\noindent The Nix package manager is the primary source of inspiration %
for Guix [8, 9].  As noted in %
\hyperref[nix-to-guix]{Section~2.3}, Guix reuses the %
low-level build and deployment mechanisms of Nix, but differs in its %
programming interface and preferred implementation language for build %
scripts.  While the Nix language relies on laziness to ensure that only %
packages needed are built [9], we instead %
support {\textit{ad hoc}} laziness with the {\texttt{package}} form.  Nix and %
Guix have the same application{\char58} packaging of a complete GNU/Linux %
distribution.\par
\noindent Before Nix, the idea of installing each package in a directory %
of its own and then managing symlinks pointing to those was already %
present in a number of systems.  In particular, the Depot [3], Store [2], and then %
GNU~Stow [10] have long supported this %
approach.  GNU's now defunct package management project called `stut', %
ca. 2005, used that approach, with Stow as a back-end.  A ``Stow file %
system'', or {\texttt{stowfs}}, has been available in the GNU~Hurd %
operating system core to offer a dynamic and more elegant approach to %
user profiles, compared to symlink trees.  The storage model of Nix/Guix %
can be thought of as a formalization of Stow's idea.\par
\noindent Like Guix and Nix, Vesta is a purely functional build system %
[11].  It uses an external DSL close to the %
Nix language.  However, the primary application of Vesta is fine-grain %
software build operations, such as compiling a single C file.  It is a %
developer tool, and does not address deployment to end-user machines. %
Unlike Guix and Nix, Vesta tries hard to support the standard Unix file %
system layout, relying on a virtual file system to ``map'' files to %
their right location in the build environment.\par
\noindent Hop defines a {\em{multi-tier}} extension of Scheme to %
program client/server web applications [17].  It %
allows client code to be introduced (``quoted'') in server code, and %
server code to be invoked from client code.  There's a parallel between %
the former and Guix's use of Scheme in two different strata, depicted in %
\hyperref[build-exprs]{Section~3.2}.\par
\noindent Scsh provides a complete interface to substitute Scheme in %
``shell programming'' tasks [18].  Since it spans %
a wide range of applications, it goes beyond the tools discussed in %
\hyperref[shell]{Section~3.4} some ways, notably by providing %
a concise {\em{process notation}} similar to that of typical Unix %
shells, and S-expression regular expressions (SREs).  However, we chose %
not to use it as its port to Guile had been unmaintained for some time, %
and Guile has since grown a rich operating system interface on top of %
which it was easy to build the few additional tools we needed.\par

\section{Conclusion}
\label{conclusion}
\noindent GNU~Guix is a contribution to package management of free %
operating systems.  It builds on the functional paradigm pioneered by %
the Nix package manager [8], and benefits from %
its unprecedented feature set---transactional upgrades and roll-back, %
per-user unprivileged package management, garbage collection, and %
referentially-transparent build processes, among others.\par
\noindent We presented Guix's two main contributions from a programming %
point of view.  First, Guix {\em{embeds}} a declarative %
domain-specific language in Scheme, allowing it to benefit from its %
associated tool set.  Embedding in a general-purpose language has %
allowed us to easily support internationalization of package %
descriptions, and to write a fast keyword search mechanism; it has also %
permitted novel features, such as an on-line auto-updater.  Second, its %
build programs and libraries are also written in Scheme, leading to a %
unified programming environment made of two strata of code.\par
\noindent We hope to make Guix a good vehicle for an innovative free %
software distribution.  The GNU system distribution we envision will %
give Scheme an important role just above the operating system %
interface.\par

\section*{Acknowledgments}
\noindent The author would like to thank the Guix contributors for their %
work growing the system{\char58} Andreas Enge, Nikita Karetnikov, Cyril %
Roelandt, and Mark H. Weaver.  We are also grateful to the Nix %
community, and in particular to Eelco Dolstra for his inspiring PhD work %
that led to Nix.  Lastly, thanks to the anonymous reviewer whose insight %
has helped improve this document.\par
\newpage
\balance %
 %

\section{References}
\label{chapter8967}
\begin{flushleft}{%
\sloppy
\sfcode`\.=1000\relax
\newdimen\bibindent
\bibindent=0em
\begin{list}{}{%
        \settowidth\labelwidth{[21]}%
        \leftmargin\labelwidth
        \advance\leftmargin\labelsep
        \advance\leftmargin\bibindent
        \itemindent -\bibindent
        \listparindent \itemindent
        \itemsep 0pt
    }%
\item[{\char91}1{\char93}] G. Beekmans,  M. Burgess,  B. Dubbs. \href{http://www.linuxfromscratch.org/lfs/}{Linux From Scratch}. 2013. \href{http://www.linuxfromscratch.org/lfs/}{{\textit{http{\char58}//www.linuxfromscratch.org/lfs/}}}.
\item[{\char91}2{\char93}] A. Christensen,  T. Egge. \href{http://dx.doi.org/10.1007/3-540-60578-9_22}{Store—a system for handling third-party applications in a heterogeneous computer environment}. Springer Berlin Heidelberg, 1995, pp. 263--276.
\item[{\char91}3{\char93}] S. N. Clark,  W. R. Nist. The Depot{\char58} A Framework for Sharing Software Installation Across Organizational and UNIX Platform Boundaries. In {\textit{In Proceedings of the Fourth Large Installation Systems Administrator’s Conference (LISA ’90)}}, pp. 37--46, 1990.
\item[{\char91}4{\char93}] R. D. Cosmo,  D. D. Ruscio,  P. Pelliccione,  A. Pierantonio,  S. Zacchiroli. \href{http://dx.doi.org/10.1016/j.scico.2010.11.001}{Supporting software evolution in component-based FOSS systems}. In {\textit{Sci. Comput. Program.}}, 76(12) , Amsterdam, The Netherlands, December 2011, pp. 1144--1160.
\item[{\char91}5{\char93}] L. Court\`{e}s. \href{http://lists.gnu.org/archive/html/bug-guix/2013-02/msg00173.html}{Boot-to-Guile!}. February 2013. \href{http://lists.gnu.org/archive/html/bug-guix/2013-02/msg00173.html}{{\textit{http{\char58}//lists.gnu.org/archive/html/bug-guix/2013-02/msg00173.html}}}.
\item[{\char91}6{\char93}] L. Court\`{e}s. \href{https://lists.gnu.org/archive/html/bug-guix/2013-01/msg00041.html}{Down with /bin/sh!}. January 2013. \href{https://lists.gnu.org/archive/html/bug-guix/2013-01/msg00041.html}{{\textit{https{\char58}//lists.gnu.org/archive/html/bug-guix/2013-01/msg00041.html}}}.
\item[{\char91}7{\char93}] O. Crameri,  R. Bianchini,  W. Zwaenepoel,  D. Kostić. Staged Deployment in Mirage, an Integrated Software Upgrade Testing and Distribution System. In {\textit{In Proceedings of the Symposium on Operating Systems Principles}}, 2007.
\item[{\char91}8{\char93}] E. Dolstra,  M. d. Jonge,  E. Visser. \href{http://nixos.org/nix/docs.html}{Nix{\char58} A Safe and Policy-Free System for Software Deployment}. In {\textit{Proceedings of the 18th Large Installation System Administration Conference (LISA '04)}}, pp. 79--92, USENIX, November 2004.
\item[{\char91}9{\char93}] E. Dolstra,  A. L\"{o}h,  N. Pierron. \href{http://nixos.org/nixos/docs.html}{NixOS{\char58} A Purely Functional Linux Distribution}. In {\textit{Journal of Functional Programming}}, (5-6) , New York, NY, USA, November 2010, pp. 577--615.
\item[{\char91}10{\char93}] B. Glickstein,  K. Hodgson. \href{http://www.gnu.org/software/stow/}{Stow---Managing the Installation of Software Packages}. 2012. \href{http://www.gnu.org/software/stow/}{{\textit{http{\char58}//www.gnu.org/software/stow/}}}.
\item[{\char91}11{\char93}] A. Heydon,  R. Levin,  Y. Yu. \href{http://doi.acm.org/10.1145/349299.349341}{Caching Function Calls Using Precise Dependencies}. In {\textit{Proceedings of the ACM SIGPLAN 2000 conference on %
Programming Language Design and Implementation}}, PLDI '00, pp. 311--320, ACM, 2000.
\item[{\char91}12{\char93}] P. Hudak. \href{http://doi.acm.org/10.1145/242224.242477}{Building domain-specific embedded languages}. In {\textit{ACM Computing Surveys}}, 28(4es) , New York, NY, USA, December 1996, .
\item[{\char91}13{\char93}] R. Kelsey. \href{http://srfi.schemers.org/srfi-9/srfi-9.html}{Defining Record Types}. 1999. \href{http://srfi.schemers.org/srfi-9/srfi-9.html}{{\textit{http{\char58}//srfi.schemers.org/srfi-9/srfi-9.html}}}.
\item[{\char91}14{\char93}] R. Kelsey,  M. Sperber. \href{http://srfi.schemers.org/srfi-35/srfi-35.html}{Conditions}. 2002. \href{http://srfi.schemers.org/srfi-35/srfi-35.html}{{\textit{http{\char58}//srfi.schemers.org/srfi-35/srfi-35.html}}}.
\item[{\char91}15{\char93}] T. Morris. \href{http://days2012.scala-lang.org/}{Asymmetric Lenses in Scala}. 2012. \href{http://days2012.scala-lang.org/}{{\textit{http{\char58}//days2012.scala-lang.org/}}}.
\item[{\char91}16{\char93}] M. Serrano,  \'{E}. Gallesio. \href{http://doi.acm.org/10.1145/1297081.1297093}{An Adaptive Package Management System for Scheme}. In {\textit{Proceedings of the 2007 Symposium on Dynamic languages}}, DLS '07, pp. 65--76, ACM, 2007.
\item[{\char91}17{\char93}] M. Serrano,  G. Berry. \href{http://doi.acm.org/10.1145/2330087.2330089}{Multitier Programming in Hop}. In {\textit{Queue}}, 10(7) , New York, NY, USA, July 2012, pp. 10{\char58}10--10{\char58}22.
\item[{\char91}18{\char93}] O. Shivers,  B. D. Carlstrom,  M. Gasbichler,  M. Sperber. \href{http://www.scsh.net/}{Scsh Reference Manual}. 2006. \href{http://www.scsh.net/}{{\textit{http{\char58}//www.scsh.net/}}}.
\item[{\char91}19{\char93}] M. Sperber,  R. K. Dybvig,  M. Flatt,  A. V. Straaten,  R. B. Findler,  J. Matthews. \href{http://journals.cambridge.org/article_S0956796809990074}{Revised6 Report on the Algorithmic Language Scheme}. In {\textit{Journal of Functional Programming}}, 19, 7 2009, pp. 1--301.
\item[{\char91}20{\char93}] R. M. Stallman. \href{http://www.gnu.org/gnu/manifesto.html}{The GNU Manifesto}. 1983. \href{http://www.gnu.org/gnu/manifesto.html}{{\textit{http{\char58}//www.gnu.org/gnu/manifesto.html}}}.
\item[{\char91}21{\char93}] D. B. Tucker,  S. Krishnamurthi. Applying Module System Research to Package Management. In {\textit{Proceedings of the Tenth International Workshop on %
Software Configuration Management}}, 2001.

\end{list}}
\end{flushleft}

\end{document}